\documentclass[prl,twocolumn,superscriptaddress]{revtex4-1}

\usepackage[T1]{fontenc}
\usepackage[latin9]{inputenc}
\usepackage{color}
\definecolor{shadecolor}{rgb}{1, 0, 0}
\usepackage{verbatim}
\usepackage{float}
\usepackage{framed}
\usepackage{amsmath}
\usepackage{amssymb}
\usepackage{graphicx}
\usepackage{comment}
\usepackage[normalem]{ulem}

\makeatletter
\PassOptionsToPackage{caption=false}{subfig}
\usepackage{hyperref}
\hypersetup{
breaklinks=true,
colorlinks=true,
citecolor=blue,
linkcolor=blue,
filecolor=blue,
urlcolor=blue
}
\IfFileExists{lmodern.sty}{\usepackage{lmodern}}{}
\makeatother

\definecolor{OliveGreen}{cmyk}{0.64,0,0.95,0.40}

\begin{document}

\title{Topological Josephson Bifurcation Amplifier: Semiclassical theory}

\author{Samuel Boutin}
\email{Sam.Boutin@gmail.com}
\affiliation{D\'epartement de physique, Institut quantique and Regroupement Qu\'eb\'ecois sur les Mat\'eriaux de Pointe,  Universit\'{e} de Sherbrooke, Sherbrooke, Qu\'{e}bec J1K 2R1, Canada}
\affiliation{Station Q, Microsoft Quantum, Santa Barbara, California 93106-6105, USA}

\author{Pedro L. S. Lopes}
\affiliation{Department of Physics and Stewart Blusson Institute for Quantum Matter, University of British Columbia, Vancouver, Canada V6T 1Z1}

\author{Anqi Mu}
\affiliation{Department of Physics, Columbia University, 538 West 120th Street, New York, New York 10027, USA}

\author{Udson C. Mendes}
\affiliation{Instituto de F\'{i}sica, Universidade Federal de Goi\'{a}s, 74.001-970, Goi\^{a}nia - Go, Brazil}

\author{Ion Garate}
\affiliation{D\'epartement de physique, Institut quantique and Regroupement Qu\'eb\'ecois sur les Mat\'eriaux de Pointe,  Universit\'{e} de Sherbrooke, Sherbrooke, Qu\'{e}bec J1K 2R1, Canada}

\date{\today}                                       
\begin{abstract}
Amplifiers based on Josephson junctions allow for a fast and noninvasive readout of superconducting qubits. Motivated by the ongoing progress toward the realization of fault-tolerant qubits based on Majorana bound states, we investigate the topological counterpart of the Josephson bifurcation amplifier. 
We predict that the bifurcation dynamics of a topological Josephson junction driven in the appropriate parameter regime may be used as an additional tool to detect the emergence of Majorana bound states.
\end{abstract}

\maketitle

\section{Introduction}

In the past decade, the discovery and characterization of Majorana bound states (MBS) in solid-state devices has become a landmark of the field of topological materials and devices \cite{reviews}. 
One promising implementation of MBS relies on semiconducting nanowires with strong spin-orbit coupling and proximity-induced superconductivity\cite{aguado2017}. In this setup, a magnetic field can induce a topological phase transition, where MBS emerge as localized states at the ends of the nanowire. Because of their spatial localization and the non-Abelian exchange statistics, these MBS are regarded as potential building blocks for fault-tolerant qubits~\cite{karzig2017}.

In anticipation of the realization of MBS-based qubits, there has been a strong interest on the theoretical front to integrate MBS in circuit quantum electrodynamics (cQED) architectures\cite{hassler2011,muller2013,pekker2013,virtanen2013,ginossar2014,yavilberg2015,dmytruk2015,vayrynen2015,peng2016,dartiailh2017,cottet2017,trif2018,keselman2019,lopes2019,avila2020,arne2019}, the latter of which are widely employed in the readout and control of solid-state qubits\cite{blais2020}.
This theoretical effort has been accompanied by experimental progress towards the realization of superconducting circuits that are compatible with sizeable magnetic fields\cite{larsen2015, delange2015, woerkom2017,luthi2018, vidal2019, tosi2019}.

Motivated by the aforementioned developments, in the present work we introduce a topological version of the Josephson bifurcation amplifier (JBA), which is simply a JBA 
made from a topological Josephson junction.
The original JBA is an amplifier designed to read out the state of superconducting qubits in topologically trivial Josephson junctions\cite{siddiqi2005, vijay}.
It is based on the transition of an RF-driven Josephson junction between two distinct oscillation states near a dynamical bifurcation point. 
The main advantages of the JBA are high speed, high sensitivity and noninvasiveness. These advantages have led to some of the first single-shot qubit readouts of superconducting qubits~\cite{mallet2009, bertet2011, schmitt2014}.
Yet, the difficulty of calibration for different devices has resulted in JBAs giving way to simpler designs for qubit measurement, such as the Josephson parametric amplifier~\cite{castellanos2007,hatridge2011,lin2013}. In our theoretical study, we show that the sensitive binary features and the high signal-to-noise ratio of the bifurcation dynamics offer new ways to detect the emergence of MBS in topological Josephson junctions. 

The remainder of this paper is organized as follows. 
Section~\ref{sec:TJBA} provides a short review of the main concepts of the JBA and introduces its topological version.
Section~\ref{sec:TPT} highlights the signatures of the topological phase transition in key parameters governing the topological Josephson bifurcation amplifier (TJBA).
In Sec.~\ref{sec:4pi}, we propose to use the TJBA in order to detect the coexisting $2\pi$ and $4\pi$ periodicities in a topological Josephson junction. 
Conclusions and final remarks are presented in Sec.~\ref{sec:conclusion}.

\section{Semiclassical theory of the TJBA}
\label{sec:TJBA}

\subsection{Josephson Bifurcation Amplifier: A review}

\begin{figure}[tb]
	\centering
    \includegraphics[width=0.8\columnwidth]{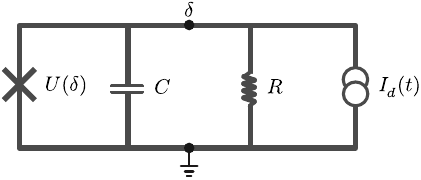}
	\caption{Circuit diagram of the RCSJ model for the dynamics of a topological Josephson junction. The cross stands for the dissipationless Cooper pair and single-particle tunneling (the latter present only in the topological phase). The AC bias current is $I_d$, while $C$ and $R$ stand for the capacitance of the junction and the resistance. 
}
	\label{fig:rcsj}
\end{figure}

We describe the topological Josephson bifurcation amplifier using the resistively and capacitively shunted (RCSJ) model of a Josephson junction\cite{likharev1986,blackburn2016}.
In this model (Fig.~\ref{fig:rcsj}), the Kirchhoff law for the electric current traversing the junction reads 
\begin{equation}
  \label{eq:rcsj}
  C \varphi_0 \frac{\partial^2\delta}{\partial t^2} + \frac{\varphi_0}{R}\frac{\partial\delta}{\partial t} + \frac{1}{\varphi_0} \frac{\partial U}{\partial\delta} = I_d(t),
\end{equation}
where $\delta$ is the gauge-invariant superconducting phase difference between the two superconducting electrodes,
$C$ and $R$ are the effective capacitance of the junction and the characteristic impedance of the microwave source generator, $\varphi_0=\hbar/2 e$ is the reduced flux quantum, $I_d$ is the bias current and $U$ is the grand potential of the junction at temperature $T$.
The first, second and third terms in the left hand side of Eq.~(\ref{eq:rcsj}) correspond to the displacement current, the dissipative ohmic current and the dissipationless (Josephson and Majorana) current, respectively. The critical current of the junction is given by $\varphi_0^{-1} {\rm max} (\partial U/\partial\delta)$.

\begin{figure*}[tb]
\includegraphics[width=\textwidth]{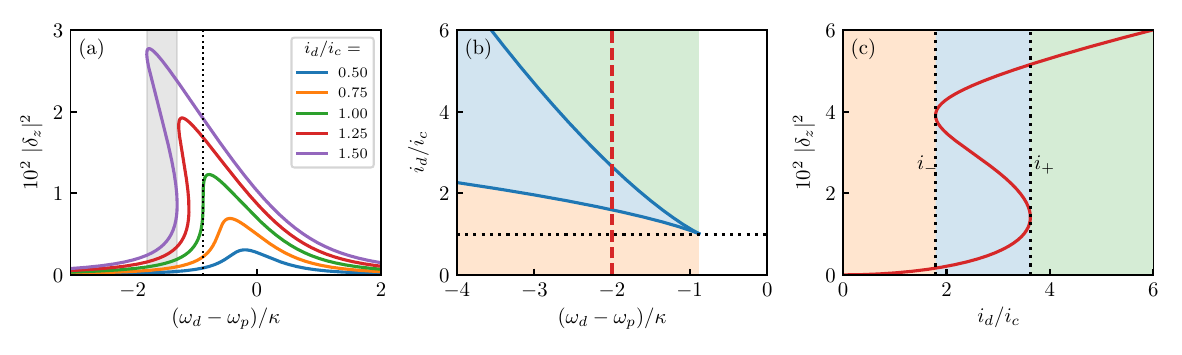}
\caption{(a) Response of a nonlinear oscillator (a Josephson junction) for different drive amplitudes $i_d$ in a frame rotating at the drive frequency $\omega_d$. Here, $\delta(t) = \mathrm{Re}[\delta_z e^{-i \omega_d t}]$ is the superconducting phase difference across the weak link, $\omega_p$ is the Josephson plasmon frequency and $i_c$ is the critical drive current for the onset of the bistable region (not to be confused with the critical current of the junction, which is parametrically higher). The vertical dotted line indicates the critical frequency of the pump, $\omega_c = \omega_p - \sqrt{3}\kappa/2$ (with $\kappa^{-1}=R C$), below which bistability emerges. The shaded area indicates the bistable regime of the $i_d=1.5 i_c$ (purple) curve.
(b) Phase diagram for bistability. The colored regions correspond to the low amplitude state (orange), the bistable regime (blue) and the high amplitude state (green). The red dashed line corresponds to panel (c). 
(c) Amplitude of $\delta$ as a function of the amplitude of the drive. From left to right, the curve is separated in three regions: (i) low amplitude state (orange), bistable regime (blue) and high amplitude state (green).
The lower and upper bifurcation currents are $i_-$ and $i_+$, respectively.}
 \label{fig:jba1}
\end{figure*}

In the RCSJ model, $\delta$ is treated as a classical variable, whereas $U(\delta)$ is computed by diagonalizing a quantum mechanical Hamiltonian for each value of $\delta$. Charging energy is not included in the Hamiltonian; instead, the phase fluctuations produced by the charging energy are incorporated via the displacement current in Eq.~(\ref{eq:rcsj}).
This semiclassical approach is justified in the "transmon regime" that is relevant to the operation of the JBA\cite{vijay}.
In such regime, the Josephson energy of the junction (approximately given by $(U(\pi)-U(0))/2$) greatly exceeds the charging energy $E_C=e^2/(2 C)$.

In thermodynamic equilibrium, fermion parity constraints are absent for timescales that are long compared to quasiparticle poisoning times. Then, the grand potential\cite{beenakker2013} takes the form
\begin{equation}
  \label{eq:UT}
  U(\delta)=-k_B T \sum_{n>0} \ln\left[2\cosh\left(\frac{\epsilon_n(\delta)}{2 k_B T}\right)\right],
\end{equation}
where $\epsilon_n(\delta)$ are the single-particle energies of the system, with $n=\pm 1, \pm 2,...$ and $\epsilon_n=-\epsilon_{-n}$ due to particle-hole symmetry (by convention, we take $\epsilon_n>0$ for $n>0$).
For $T\rightarrow 0$, Eq.~(\ref{eq:UT}) reduces to the ground state energy
\begin{equation}
  \label{eq:U_T0}
  U(\delta)=-\frac{1}{2}\sum_{n>0} \epsilon_n(\delta).
\end{equation}
The eigenvalues $\epsilon_n(\delta)$ can be obtained by diagonalizing a tight-binding Hamiltonian describing a single-channel, one-dimensional superconducting/normal/superconducting heterostructure of finite length, where two long superconducting electrodes are separated by a weak link in the normal state.
More details of the model are presented in Sec.~\ref{ssec:model}.

Equation~\eqref{eq:rcsj} describes the dynamics of a fictitious particle of mass $C$ and position $\varphi_0\delta$ 
moving in a potential $U(\delta)$ under the action of an external force $I_d$.
We consider a noisy sinusoidal AC-bias current,
\begin{equation}
  I_d(t) = i_d \sin(\omega_d t) + I_N(t),
\end{equation}
where $\omega_d$ and $i_d$ are, respectively, the drive frequency and amplitude, and $I_N(t)$ is a white-noise current produced by thermal fluctuations.
We will focus in the case in which $i_d$ is small compared to the critical current of the junction. 
Then, the dynamics of the particle is akin to that of a nonlinear harmonic oscillator centered in one of the minima of $U(\delta)$, which are 
located at $\delta_m=2\pi m$ ($m\in\mathbb{Z}$).
Because $U(\delta)=U(\delta+2\pi)$ in the absence of parity constraints,  we can concentrate on oscillations near $\delta=0$ without loss of generality. 

Following the small oscillation approximation, Eq.~\eqref{eq:rcsj} can be rewritten as 
\begin{equation}
\label{eq:rcsj2}
   \frac{\partial^2\delta}{\partial t^2} + \kappa \frac{\partial\delta}{\partial t} + 
   \omega_p^2 \delta + \lambda \delta^3
   = \frac{I_d(t)}{C \varphi_0},
\end{equation}
where the damping rate $\kappa =1/RC$, the plasma frequency $\omega_p$ in the harmonic approximation and the leading anharmonicity parameter $\lambda$ play a central role in the functioning of the JBA.
They can be extracted from $U$ via
\begin{align}
\begin{split}
  \label{eq:jba_par}
  \omega_p &= \sqrt{8 U^{(2)}(0) E_C}/\hbar \\
  \lambda &= \frac{4 U^{(4)}(0)E_C}{3 \hbar^2},
  \end{split}
\end{align}
where $U^{(n)}(0) = (\partial^n U/\partial\delta^n)|_{\delta=0}$. 

Figure~\ref{fig:jba1} illustrates its solution in the absence of current noise.
The various curves in this figure are well-known in the literature\cite{vijay}; here, we highlight the most important points for completeness and later reference.
In a frame rotating at the drive frequency, i.e. taking $\delta(t) = \mathrm{Re}[\delta_z e^{-i \omega_d t}]$, we solve for the steady-state solution of Eq.~\eqref{eq:rcsj2}.
Ignoring rapidly oscillating terms (rotating-wave approximation), Fig.~\ref{fig:jba1}a displays the modulus of $\delta_z$ as a function of $\omega_d$, for different values of $i_d$. 
For small $i_d$, the response of the oscillator is a Lorentzian peaked at $\omega_d=\omega_p$.
As the driving amplitude increases, the response of the oscillator becomes peaked at frequencies lower than $\omega_p$ (this is due to the fact that the anharmonicity parameter in Eq.~(\ref{eq:jba_par}) is negative).
Beyond a critical value of $i_d$ (but still well below the critical current of the junction), there emerges a finite interval of $\omega_d$ for which the response of the oscillator is multi-valued, with three values of $|\delta_z|$ for each $\omega_d$.
In such an interval, the intermediate value of $|\delta_z|$ is not stable and the oscillator is thus said to be in a bistable regime, with a low- and a large-amplitude oscillation steady state.

Figure \ref{fig:jba1}b shows the dependence of the bistable regime on the drive frequency.
Bistability takes place when $\omega_d\leq \omega_c\equiv \omega_p - \sqrt{3} \kappa/2$ 
and $i_d\in(i_-, i_+)$, where $i_-$ and $i_+$ are the lower and upper bifurcation currents, respectively.
These currents obey the relations
\begin{align}
\begin{split}
  \label{eq:i_pm}
  i_\pm^2 &= 
  -\frac{8 \varphi_0^2 C^2}{81 \lambda} 
  \left[2\omega_\delta \omega_\Sigma \pm \sqrt{r(\omega_d)}\right]
    \\
  &\times\left[3 \omega_d^2 \kappa^2 + \omega_\delta^2 \omega_\Sigma^2 \mp \omega_\delta \omega_\Sigma \sqrt{r(\omega_d)}\right],
  \end{split}
\end{align}
where $r(\omega_d)=(\omega_d^2-\omega_p^2)^2-3 \omega_d^2\kappa^2$, $\omega_\delta=\omega_p-\omega_d$ and $\omega_\Sigma=\omega_d+\omega_p$.
At the bifurcation currents, the response of the oscillator switches between single-valued and two-valued.
Thus, for $i_d<i_-$ ($i_d>i_+$), the oscillation amplitude is single-valued and low (high).
The fact that bifurcation currents  depend on $\omega_p$ and $\lambda$ will be important below.

Figure~\ref{fig:jba1}c presents the dependence of the steady-state solution $|\delta_z|$ on the drive amplitude, for a fixed value of the drive frequency ($\omega_d<\omega_c$, red dashed vertical line in Fig.~\ref{fig:jba1}b). In the bistable regime (blue shaded central region), which of the two stable steady-states is reached depends on the initial state of the system and how the drive is turned on. In the case where the drive is turned on sufficiently slowly (compared to the timescale $1/\kappa$) and in the absence of current noise, the steady state follows an hysteresis curve. For an initial state where $|\delta_z| \sim 0$,  the low-amplitude steady-state is realized for any $i_d<i_+$, and the system suddenly jumps to the high-amplitude solution at $i_d=i_+$.
Driving the the Josephson junction close to $i_+$ thus allows to detect small changes in the upper bifurcation current, 
independently from the width of the bistable region $|i_+-i_-|$.

\begin{figure}[tb]
\includegraphics[width=0.78\columnwidth]{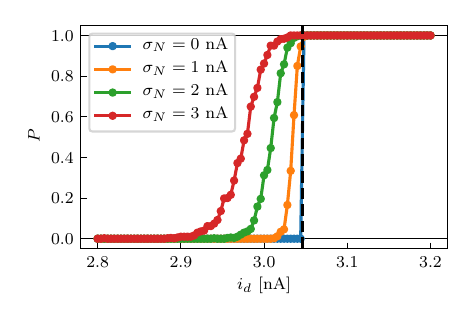}
\caption{Probability of finding the Josephson oscillator in the high-amplitude state for different noise strengths after evolution time $T_0$. In order to extract the probability, Eq.~(\ref{eq:rcsj2}) is solved in the rotating frame for $M=500$ noise realizations. In all cases, we take the initial condition $\delta(0)=\partial_t\delta(0)=0$,  and the parameters $\kappa=\omega_p/25$, $\omega_d=\omega_p-3\kappa/2$, $\omega_p/2\pi = 5$~GHz, $E_C/h = 0.1$~GHz and $\kappa T_0\sim 250$. Due to the hysteretic nature of the JBA, in the absence of noise the transition of $P$ from $0$ to $1$ takes place at $i_d=i_+$ (dashed black line) for a smooth switch-on of the drive on a timescale $\tau$ with $\kappa \tau \gg 1$ (here $\kappa \tau\sim 30$).
}
 \label{fig:scurve}
\end{figure}

Thus far we have reviewed the phenomena of bifurcation and bistability in the absence of current noise ($I_N=0$).
In the presence of noise, due to either thermal or quantum fluctuations, the steady-state reached by the system is not deterministic. We define the bifurcation probability $P$ as the probability of the system reaching the high-amplitude state  after evolution of the system for a time $T_0$~\cite{vijay,bertet2011}. 
To estimate $P$, we integrate Eq.~(\ref{eq:rcsj2}) for a time $T_0>1/\kappa$ and $M\gg 1$ realizations of the noise $I_N(t)$.
We consider uncorrelated white noise by taking $I_N(t)$ from a zero-mean Gaussian distribution of variance $\sigma_N^2= 4 k_B T/R$.
We then estimate the bifurcation probability by
$P \approx n_H/M$,
where $n_H$ is the number of high amplitude steady-state solutions.
This probability $P$ is experimentally measurable. 

Figure~\ref{fig:scurve} illustrates the dependence of $P$ on the drive amplitude in the presence of current noise.
When $i_d < i_-$ and $i_d> i_+$, the oscillator is with certainty in the low and high amplitude states, respectively.
Due to the hysteretic nature of the bifurcation dynamics, $P$ is a sharp step function at $i_d=i_+$ in the absence of noise (blue curve). Current noise increases the width of the step function in the region $i_d \lesssim i_+$. As we consider a low amplitude initial state and a smooth turn on of the drive, $P$ does not depend on the lower bifurcation current ($i_-\approx 2.25$~nA in Fig.~\ref{fig:scurve}). 
Because the bifurcation currents are sensitive to the intrinsic parameters of the junction (such as $\omega_p$ and $\lambda$), small variations of the latter lead to significant changes in $P$ when $i_d$ lies in the vicinity of $i_+$.
As we explain below, this high sensitivity may allow for new ways to measure MBS properties in topological Josephson junctions.

\begin{figure}[tb]
	\centering
	\includegraphics[width=0.8\columnwidth]{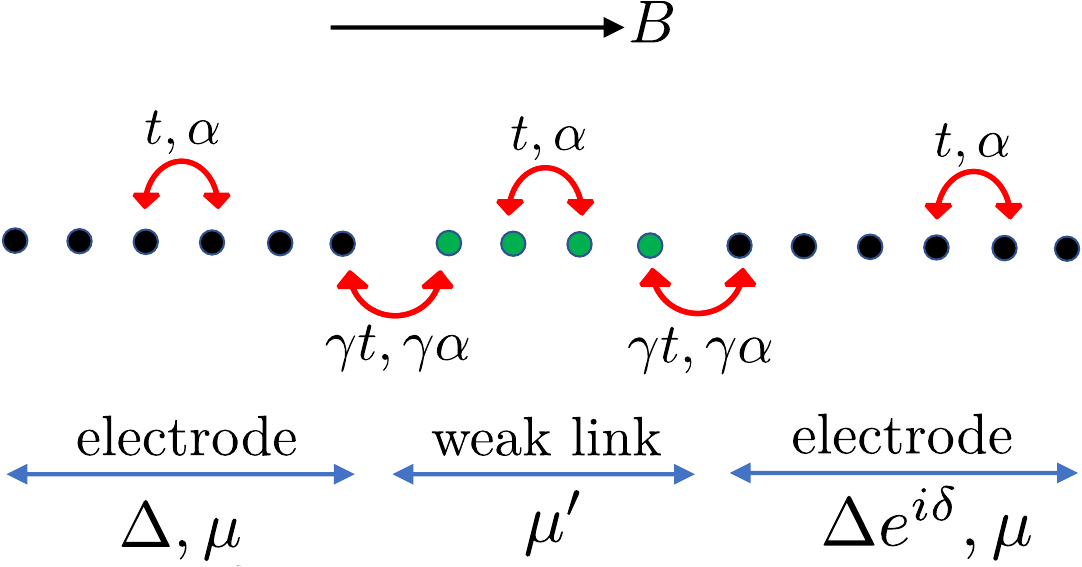}
	\caption{Pictorial representation of the tight-binding model for the topological Josephson junction.
	Green dots correspond to the weak link. The spin-independent and spin-flip tunneling amplitudes are $t$ and $\alpha$, respectively. The parameter for the interfaces between the superconducting electrodes and the weak link is $\gamma\leq 1$. The proximity-induced $s$-wave superconducting gap on the electrodes has an amplitude $\Delta$, with a phase difference $\delta$ between the two electrodes. The Zeeman splitting caused by the external magnetic field is $B$. The chemical potential in the electrodes is $\mu$, while that of the weak link is $\mu'$. Unless otherwise stated, we take $\mu=\mu'$.}
	\label{fig:tb}
\end{figure}

\subsection{Limiting forms of $U(\delta)$}\label{ssec:limits}

Before turning to a more microscopic description of a Josephson junction by calculating the grand potential $U(\delta)$ from a tight-binding model, it is helpful to consider two approximate limiting forms of $U(\delta)$ in ideal tunnel junctions. These limiting cases will be useful to interpret some of the numerical results of the following sections.
%

In the case of conventional tunnel junction at low temperature, the ground state energy is $U(\delta)=-E_J \cos\delta+{\rm const}$, where $E_J$ is the Josephson energy associated to the tunneling of Cooper pairs and "const" is a $\delta-$independent number. 
From Eq.~\eqref{eq:jba_par}, the plasmon frequency and anharmonicity parameter are then $\hbar \omega_p = \sqrt{8 E_J E_C}$ and $\lambda = -\omega_p^2/6$. 
In this case, $\lambda/\omega_p^2$ is a constant because $U(\delta)$ has a single harmonic ($\cos\delta$).

In the case of a topological tunnel junction, one instead finds (disregarding parity constraints) $U(\delta)\simeq -E_M |\cos(\delta/2)|-E_J \cos\delta +{\rm const}$, where $E_M$ is the Majorana energy associated to the tunneling of single electrons. 
In particular, if  the junction transparency is very low, $E_J\ll E_M$ and $U(\delta)\simeq -E_M |\cos(\delta/2)|$.
In this tunnel regime where Cooper pair tunneling across the junction is negligible compared to single-particle tunneling, we have
$\omega_p\sim \sqrt{2 E_M E_C}/\hbar$ and $\lambda\sim-\omega_p^2/24$.
Once again, $\lambda/\omega_p^2$ is a negative constant, but of smaller amplitude. 

In general, when the $\delta-$dependence of $U(\delta)$ has more than one harmonic (e.g. in topological junctions with comparable $E_J$ and $E_M$, or even in trivial junctions that are neither in the tunneling nor in the perfectly transparent regime\cite{golubov2004}), $\lambda/\omega_p^2$ will not be a constant.

Finally, for later reference,
we note that in the literature of superconducting qubits\cite{blais2020}, anharmonicity in the transmon regime is often defined as $\lambda'=(E_2-E_1)-(E_1-E_0)$, where $E_n$ are the energy levels for the quantum anharmonic oscillator evaluated to first order in the strength of the quartic potential. The relation between $\lambda'$ and our anharmonicity parameter is $\lambda'= 6 \lambda E_C/\omega_p^2$. Thus, $\lambda'=-E_C$ in a conventional tunnel junction.

\subsection{Microscopic model of a topological junction}\label{ssec:model}
In order to account for topological effects,
we calculate the grand potential $U(\delta)$ defined in Eq.~\eqref{eq:UT} by diagonalizing a tight-binding model of a 1D nanowire junction (see Fig.~\ref{fig:tb} for a sketch of the model).
As this model has been extensively reviewed in the literature, see e.g. Ref.~[\onlinecite{cayao2018}], we limit ourselves to providing a brief description.

We consider spinful fermions with  annihilation (creation) operator $c_{j, \sigma}^{(\dag)}$ ($\sigma \in \left\{\uparrow, \downarrow\right\}$). Introducing the spinor $\psi^\dag_j = ( c_{j,\uparrow}^\dag, c_{j,\downarrow}^\dag, -c_{j,\downarrow}, c_{j,\uparrow})$, the Hamiltonian reads
\begin{align}
    H = \sum_j \psi^\dag_j h_{i}\psi_{j} + \left(\psi^\dag_{j+1} u_{j}\psi_{j} + h.c.\right),
\end{align}
where $h_j$ and $u_j$ are (respectively) the onsite and hopping matrices, with site-dependent parameters in order to create a junction. We define these matrices in terms of two sets of Pauli matrices, $\sigma_{\alpha}$ and $\tau_{\alpha}$ ($\alpha \in \left\{ x,y,z\right\})$, acting respectively on the spin and the particle-hole sector of the spinor.
Electrons can hop between lattice sites with 
a spin-independent nearest-neighbor tunneling amplitude $t$ and,  due to spin-orbit coupling, a spin-flip tunneling amplitude $\alpha$.
Then, the hopping matrix is
\begin{align}
    u_j = -t_j \tau_z -i \alpha \sigma_y \tau_z,
\end{align}
where $t_j=t$ away from the junction interface (see Fig.~\ref{fig:tb} and below).
The onsite matrix reads
\begin{align}
    h_j = (2t-\mu)\tau_z +B\sigma_x +  \Delta_j \tau_x,
\end{align}
where the local mean-field superconducting gap $\Delta_j$ is 
$\Delta$ in the left superconducting lead, 0 in the weak link and $\Delta e^{i\delta}$ in the  right superconducting lead.
The Zeeman energy $B$ associated to the applied magnetic field is assumed to be spatially homogeneous.
Likewise, the chemical potential $\mu$ is assumed to be uniform and identical in the two superconducting electrodes, though we allow for a mismatch in the chemical potentials of the electrodes and the weak link. The value $\mu=0$ corresponds to the chemical potential being in the middle of the Zeeman gap of the normal state electronic bands.

Using parameters relevant to InAs nanowires~\cite{aguado2017}, we take $t=25$~meV and $\alpha=t/10 \sim 2.5$~meV (corresponding to a lattice constant $a=10$~nm and 
an electronic effective mass $m^*=0.016 m_e$, with $m_e$ the bare electron mass). 
We take throughout the proximity-induced $s$-wave superconducting gap to be $\Delta=0.9$~meV. 
We also neglect the dependence of $\Delta$ on $B$ and on $\mu$. 
In reality, this dependence is smooth across a topological phase transition, and thus clearly distinguishable from the sharp features we will be discussing below.
It is well-known that variations of $B$ and $\mu$ can drive the superconducting electrodes across a topological phase transition\cite{reviews2}.
The superconducting electrodes are in the topological phase (with a MBS in each extremity of the superconductors) for $B>B_c \equiv \sqrt{\Delta^2+\mu^2}$, and in the trivial phase otherwise. 

To model the effect of a potential barrier at the superconductor/normal interface, we use a phenomenological transparency parameter $\gamma\in[0,1]$. Both the hopping amplitude $t$ and the spin-orbit coupling parameter $\alpha$ are scaled by $\gamma$ at the interface.
Throughout the text, we consider the cases $\gamma=0.6$ and $\gamma=1$ to study respectively the effect of a large potential barrier at the junction interface and the case of a clean junction with no potential barrier. We denote the former as the "tunnel" regime and the latter as the "transparent" regime. 
However, we note that the junction transparency also depends on the normal state Fermi velocity $v_F$~\cite{BTK, heck2018} and is hence a function of $B$, $\alpha$ and most importantly $\mu$. For example, even when $\gamma=1$, the junction's characteristics (e.g. $\omega_p$) can resemble those of a true tunnel junction if $\mu$ is sufficiently negative.

In the numerical simulations, unless mentioned otherwise, 
the weak link is two sites long ("short junction" regime), whereas each superconducting electrode contains $200$ sites.
The main qualitative points are relatively insensitive to the strength of spin-orbit coupling, though larger values of $\alpha$ make it easier to observe the relevant features.
Likewise, longer junction lengths do not change the main features discussed below, though they host additional properties that complicate the JBA-based detection of the topological phase transition; we elaborate on this issue below.

\begin{figure*}[t]
    \includegraphics[width=0.99\textwidth]{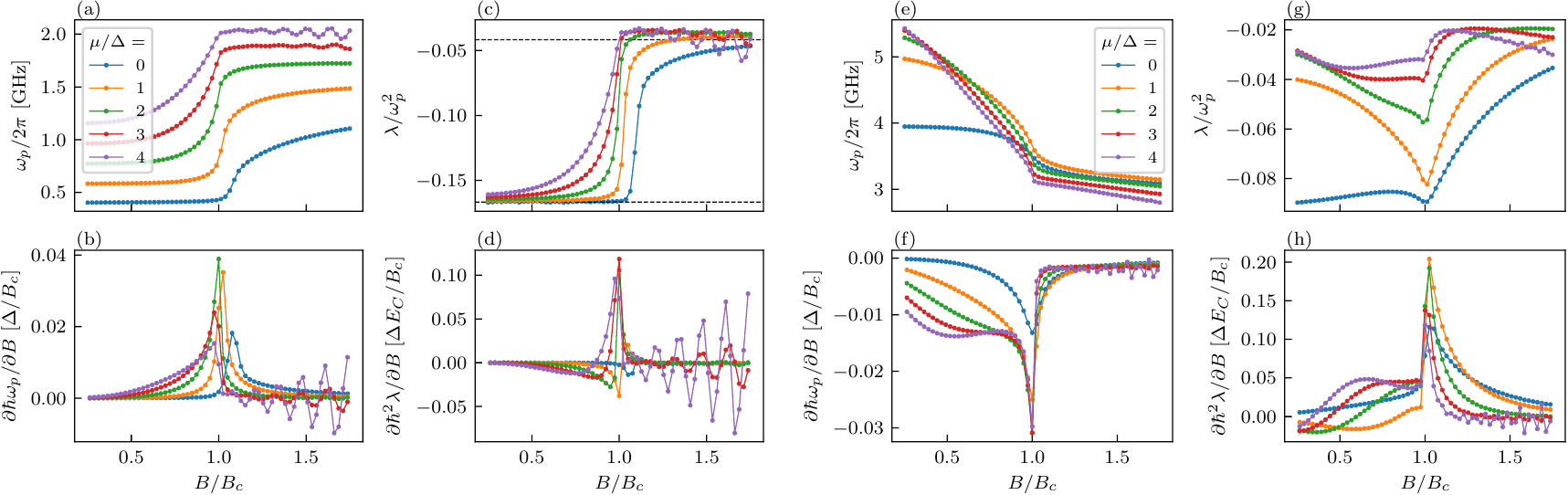}
\caption{
Dependence of the Josephson plasmon frequency $\omega_p$ and the anharmonicity parameter $\lambda$ on the Zeeman energy $B$, for a short junction at zero temperature. We take a small charging energy $E_C/h =100$~MHz; see main text for other parameters of the model. The derivatives of $\omega_p$ and $\lambda$ with respect to $B$ are also shown as a function of $B$. In the thermodynamic limit, the topological phase transition takes place at $B_c=\sqrt{\Delta^2+\mu^2}$, $B>B_c$ corresponding to the topological regime. (a-d) "Tunneling" regime ($\gamma=0.6$), (e-h) "Transparent" regime ($\gamma=1$). (c) Dashed lines indicate the expected value of $\lambda/\omega_p^2=-1/6$ for a ground state energy of the form $U(\delta) \propto \cos \delta$, and $\lambda/\omega_p^2=-1/24$ for $U(\delta) \propto \cos (\delta/2)$.
}
 \label{fig:B1}
\end{figure*}
\section{Detection of the topological phase transition}
\label{sec:TPT}

In the preceding section, we have demonstrated that the JBA can be highly sensitive to changes in the current-phase characteristics of a Josephson junction. It is therefore natural to wonder whether a JBA could detect the emergence of Majorana bound states. 
Here, we investigate the signatures of the topological phase transition in key JBA observables.
We begin by discussing the plasmon frequency and the anharmonicity parameter at zero (Sec.~\ref{ssec:wpT0}) and finite (Sec.~\ref{ssec:wpFinitT}) temperature. This analysis sets the stage for the subsequent computation (Sec.~\ref{ssec:bif}) of bifurcation currents across the topological phase transition. In this section, we will focus on thermodynamic equilibrium, where Eq.~(\ref{eq:UT}) applies.
In Sec.~\ref{sec:4pi}, we will depart from thermodynamic equilibrium in order to calculate $\omega_p$ and $\lambda$ for different many-body states.

\subsection{Plasmon frequency and anharmonicity parameter at zero temperature}\label{ssec:wpT0}

Figure \ref{fig:B1} shows the dependence of the Josephson plasmon frequency $\omega_p$ and the anharmonicity parameter $\lambda$ 
on the magnetic field, for small oscillations of the superconducting phase difference in the vicinity of $\delta=0$, in the many-body ground state.
We show the results both in the "tunneling" regime ($\gamma=0.6$, Fig.~\ref{fig:B1}a-d) and in the "transparent" regime ($\gamma=1$, Fig.~\ref{fig:B1}e-h).
In all cases, the most salient feature is the pronounced peak in  $\partial \omega_p/\partial B$ and $\partial\lambda/\partial B$ at the topological phase transition.
As we explain next, these peaks can be attributed to the emergence of MBS.

In the tunneling regime\cite{tunnel1}, Fig.~\ref{fig:B1}a evidences an abrupt growth of the plasmon frequency at the onset of the topological phase ($\omega_p$ grows by more than $0.1$~GHz as $B/B_c$ changes by less than $10\%$).
This behavior can be understood by analyzing $\epsilon_n(\delta)$ as a function of $\delta$. In our model for a topologically trivial tunnel junction, all single-particle "bands" are rather flat in $\delta$ (they would be completely independent of $\delta$ if $\gamma=0$). 
Accordingly, $U(\delta)$ depends weakly on $\delta$ and hence $\omega_p$ is relatively small.
In the topological phase, the single-particle bands remain flat, except for those that result from the hybridization of MBS. The latter disperse parametrically more strongly with $\delta$, because their "bandwidth" scales with the square root of the transparency, rather than with the transparency itself (as is the case for the non-Majorana bands). 
Hence, this explains why $\omega_p$ increases strongly at the onset of the topological phase, thus producing a strong peak in $\partial \omega_p/\partial B$ (Fig.~\ref{fig:B1}b). 

The aforementioned argument, derived in the context of a single-channel nanowire model, applies also to a less idealized tunnel junction that may host multiple subbands in the normal state. 
The simplest way to see this is by adding a term of the form $-E_{J0} \cos\delta$ to the ground state energy, where $E_{J0}$ models the contribution from the multiple subbands to the conventional Josephson energy of the junction. 
As a result of the additional term in the energy, $\omega_p$ is shifted upwards in both the trivial and the topological regimes, but the ramp up of $\omega_p$ and the peak of $\partial \omega_p/\partial B$ at the onset of the topological phase remain unchanged\cite{EJ0}.

A related behavior has been identified for the DC critical current in earlier works, with a marked increase at the onset of the topological phase \cite{sanjose2014, cayao2017, huang2017}.
At first glance, a similarity in the behavior of $\omega_p$ and the critical current is not surprising, the latter being proportional to $\omega_p^2$ in conventional Josephson junctions.
Yet, for the unconventional junctions we are interested in, this relation of proportionality does not apply. Instead, all one can assert is that $\omega_p$ depends on the characteristics of the single-particle bands near $\delta=0$, while the critical current depends on the slope of the single-particle bands near large ($\simeq \pi$) values of $\delta$.
Moreover, from an experimental point of view, the measurement of $\omega_p$ is qualitatively different from the measurement of the critical current; it can therefore offer an alternative way to probe the interpretation of a recent experiment reporting an enhanced critical current at the onset of the topological phase\cite{tiira2017}.

Let us now discuss the anharmonicity parameter. Still in the tunneling regime, Fig.~\ref{fig:B1}c reflects a peculiar evolution of $\lambda/\omega_p^2$ across the topological transition, which can be understood analytically. 
Deep in the trivial regime, we find $\lambda/\omega_p^2=-1/6$, as expected for a conventional Josephson junction with $U(\delta)=-E_J \cos\delta$.
Deep in the topological regime, $\lambda/\omega_p^2$ tends asymptotically to $-1/24$, as expected for $U(\delta)=-E_M |\cos(\delta/2)|$ 
(cf. Sec.~\ref{ssec:limits}).
In other words, in a tunnel junction, the evolution of $\lambda/\omega_p^2$ across the topological phase transition reveals the emergence of MBS. 

Figure~\ref{fig:B1} also displays the results for several values of $\mu/\Delta$.
Even though we consider $\mu$ to be homogeneous throughout the system (i.e., $\mu=\mu'$ in Fig.~\ref{fig:rcsj}b), we have verified that the results do not change significantly when we vary $\mu$ in the superconducting electrodes while keeping $\mu'$ pinned to zero in the weak link. 
We note that $\omega_p$ increases with $\mu/\Delta$ in Fig.~\ref{fig:B1}.
This effect can be understood by observing that, for fixed $\gamma$,  the junction transparency is a monotonically decreasing function of the ratio of the barrier amplitude and the normal-state Fermi velocity $v_F$~\cite{BTK}.
As $\mu/t \ll 1$ for all the cases considered (Fermi level near the bottom of the band), increasing $\mu$ means a higher $v_F$ and hence an increase of the junction transparency.
Although Fig.~\ref{fig:B1} shows only positive values of $\mu$, we have checked that, for $\mu<0$, the junction transparency effectively decreases. For $\mu/\Delta\lesssim -4$, the behavior expected for a tunnel junction is found irrespective of the value of $\gamma$ as the Fermi level in the normal state then lies below the bottom of the band (insulating regime).

Another effect of increasing $\mu/\Delta$ is the appearance of oscillations in $\omega_p$ and $\lambda$ as a function of $B$; their amplitude grows with $B$.
These oscillations are the microwave counterpart of the ones predicted in the DC critical current by Cayao {\em et al.}\cite{cayao2017}
The oscillations take place only in the topological phase and arise from the hybridization between MBS localized at opposite extremities of each superconducting electrode, as well as from the hybridization between the MBS localized at the two opposite extremities of the entire system.
Accordingly, the oscillations are washed out when the length of the superconducting electrodes becomes long enough (we find no evidence of them when doubling the electrodes size to $400$ sites). 

Let us now discuss the transparent junction regime $\gamma=1$ (see Fig.~\ref{fig:B1}e-h). 
In this regime, the behavior of $\omega_p$ at the topological phase transition is also reminiscent of that predicted in earlier works for the DC critical current.\cite{sanjose2013, cayao2017}
In those works, a kink of the critical current at the topological phase transition was identified and attributed to a band inversion taking place at the topological phase transition.
A similar mechanism is at play in our case, as we explain next.

The underlying explanation begins by recognizing that a long superconducting electrode in the trivial phase hosts two different and quasi-independent $p$-wave superconducting gaps in the energy spectrum\cite{sanjose2013, cayao2017, murthy2020}: one at the inner Fermi points (the "$\Delta_-$ gap") and the other at the outer Fermi points (the "$\Delta_+$ gap").
When $B\ll \alpha$, $\Delta_+\simeq \Delta$.
As $B$ increases, $\Delta_+$ decreases gradually and gets significantly suppressed when $B\gg \alpha$.
The $\Delta_-$ gap is far more sensitive to $B$, vanishing when $B=B_c$.
This gap-closing point marks the topological phase transition.
Near the transition, we have $\Delta_-\simeq |B-B_c|$.

The next part of the explanation is to recall that in a short, transparent and topologically trivial junction, there is a single Andreev bound (or quasi-bound) state associated to each of the two $p$-wave gaps of the electrodes\cite{tunnel}.
In this regime, the rest of the single-particle states (the so-called scattering states) are largely dispersionless in $\delta$, and therefore the main contribution to $\omega_p$ and $\lambda$ originates from the two Andreev states bound by $\Delta_+$ and $\Delta_-$.
Although there is no simple analytical expression for the Andreev bound state dispersions in the presence of a generic magnetic field\cite{heck2018}, qualitative insight can be gained by neglecting the coupling between the two $p$-wave gaps, adopting the Andreev approximation, and assuming perfect transparency. Then, we posit
$\epsilon_\pm(\delta) \simeq \Delta_\pm \sqrt{1-\sin^2(\delta/2)}$ for the two ABS in the trivial phase, and we get
\begin{align}
\label{eq:analytical}
      \omega_p &\simeq  \sqrt{(\Delta_-+\Delta_+) E_C}/\hbar\nonumber\\
       \lambda &\simeq -(\Delta_-+\Delta_+) E_C/ (24 \hbar^2).
\end{align}

Across a topological phase transition from the trivial to the topological phase, $\Delta_-$ crosses zero and inverts its sign, thereby changing its character from a superconducting gap to a magnetic gap. 
Accordingly, in the topological phase, $\Delta_-$ no longer binds an Andreev state. 
Therefore, the contribution from $\Delta_-$ to $\omega_p$ and $\lambda$ in Eq.~(\ref{eq:analytical}) is turned off for $B\geq B_c$.
This results in a kink for $\omega_p$ and $\lambda$ as a function of $B$ at $B=B_c$ (and corresponding discontinuities in $\partial\omega_p/\partial B$ and $\partial \lambda/\partial B$). Of course, such reasoning of "suddenly turning off" the contribution from $\Delta_-$ at $B=B_c$ makes sense only at zero temperature; finite-temperature effects will be discussed in Sec.~\ref{ssec:wpFinitT}.

Some of the features in Fig.~\ref{fig:B1} match qualitatively with the behavior expected from Eq.~(\ref{eq:analytical}).
First, $\omega_p$ is decreasing as we approach the topological phase transition. Second, a kink is visible at the transition, especially for larger values of $\mu/\Delta$, and $\omega_p$ continues to decrease as we get deeper in the topological phase (because $\Delta_+$ keeps decreasing as $B$ is made larger). 
On the other hand, the results from Fig.~\ref{fig:B1}g do not match qualitatively with Eq.~(\ref{eq:analytical}).
For one thing, $\lambda/\omega_p^2$ is not a constant. 
Also, a clear kink in $\omega_p$ at the topological phase transition is present only if $\mu/\Delta$ is sufficiently large.
There are various possible reasons for these discrepancies. 

First, the analytical expressions for $\epsilon_\pm(\delta)$ are approximately valid at $B=0$, but may fail qualitatively at larger $B$. 
Even at $B=0$, $\gamma=1$ does not guarantee a perfect transparency of the junction; the actual transparency depends on the value of $\mu$ as well. 
If the transparency is not unity, one expects various competing harmonics in $\epsilon_\pm (\delta)$ near $\delta=0$ (rather than just $\cos(\delta/2)$), which in turn leads to $\lambda/\omega_p^2$ not being constant.
Second, the Andreev approximation upon which Eq.~(\ref{eq:analytical}) is based is not satisfied in the low electron density regime required to have a topological Josephson junction; this is evidenced by the single-particle energy spectra we have calculated (not shown), wherein the ABS in the trivial phase have significant gaps at $\delta=\pi$.
Third, near $\delta=0$, the Andreev state bound by $\Delta_+$ is buried in the continuum of scattering states\cite{sanjose2013, murthy2020}; this is due to the fact that $\Delta_+\gg\Delta_-$ in the vicinity of the phase transition.
In other words, the ABS bound by $\Delta_+$ undergoes multiple avoided crossings with scattering states near $\delta=0$.
Consequently, it is not accurate to neglect the $\delta$-dependence of those scattering states, nor is it to ignore their contribution to $\omega_p$ and $\lambda$.

Next, we compare our results to relevant works in the literature. Our finding that $\partial\omega_p/\partial B $ and $\partial \lambda/\partial B$ display prominent peaks at $B=B_c$ is seemingly more
optimistic than that of a recent numerical study by Keselman {\em et al.}\cite{keselman2019}, who conclude that there are "no strong signatures of the topological transition itself on the simulated frequency spectra of the junction."
This discrepancy in viewpoint might be explained by the different parameter regimes considered. In particular, the exact numerical treatment of charging energy in Ref.~[\onlinecite{keselman2019}] comes at the expense of studying smaller systems ($\sim 40$ sites), where finite-size effects smoothen the phase transition into a crossover. Below, we show that finite-temperature effects can likewise broaden and eventually erase the sharp signatures in $\partial \omega_p /\partial B$.
In another recent work\cite{avila2020}, Avila {\em et al.} have, to the contrary, reported an abrupt dip in the anharmonicity parameter $\lambda'$ at $B=B_c$, stating it to be a "precise smoking gun" for the topological phase transition.
We recall (see discussion at the end of Sec.~\ref{ssec:limits}) that $\lambda'= 6 E_C \lambda/\omega_p^2$.

The theoretical treatments of Avila {\em et al.} and Keselman {\em et al.} differ from ours in that they treat the charging energy quantum mechanically, while we do so semiclassically.
Yet, both approaches ought to yield similar results in the transmon regime. In Fig.~\ref{fig:B1}g, we do find that $\lambda/\omega_p^2$ can have a dip at $B=B_c$ in the transparent regime, though it is nowhere as pronounced as in Fig. 9b of Ref.~[\onlinecite{avila2020}].
Moreover, the dip is not a generic feature when we vary parameters such as $\gamma$ and $\mu$ (e.g., it is absent in Fig.~\ref{fig:B1}c).
What appears more generic in our case is the peak in $\partial\lambda/\partial B$ at $B=B_c$, whose origin we have explained above.

On a related note, while Avila {\em et al.} investigate the behavior of the nonlinear Josephson inductance in the topological and trivial phases, they do not point out specific signatures at the phase transition (at least Fig. 8 of Ref. [\onlinecite{avila2020}] does not show enough values of $B$ to ascertain any signatures). 
Yet, the inverse of this nonlinear inductance, close to $\delta=0$, is proportional to $\omega_p^2$. 
It would be interesting to find out whether the theory of Ref.~[\onlinecite{avila2020}] agrees with our prediction for  $\partial\omega_p/\partial B$ near $B_c$.

\begin{figure}[t]
  \includegraphics[width=0.9\columnwidth]{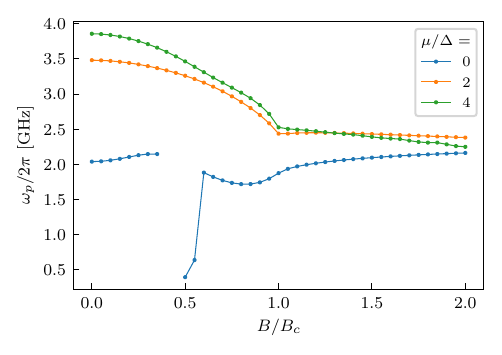}
\caption{Behavior of the plasmon frequency $\omega_p$ as a function of the Zeeman energy $B$, for 20 sites in the weak link, high transparency ($\gamma=1$) and zero temperature. A discontinuity is present in the trivial phase ($B<B_c$) for the case $\mu=0$. This discontinuity emerges from the accidental zero-energy crossing of low-lying Andreev bound states at $\delta=0$. No such crossing is present at the topological phase transition ($B=B_c$), where low-lying Andreev-bound states instead coalesce to zero energy.
Thus, there are no discontinuities in $\omega_p$ at $B=B_c$.}
 \label{fig:long_disc}
\end{figure}
\begin{figure*}[tb]
    \includegraphics[width=0.99\textwidth]{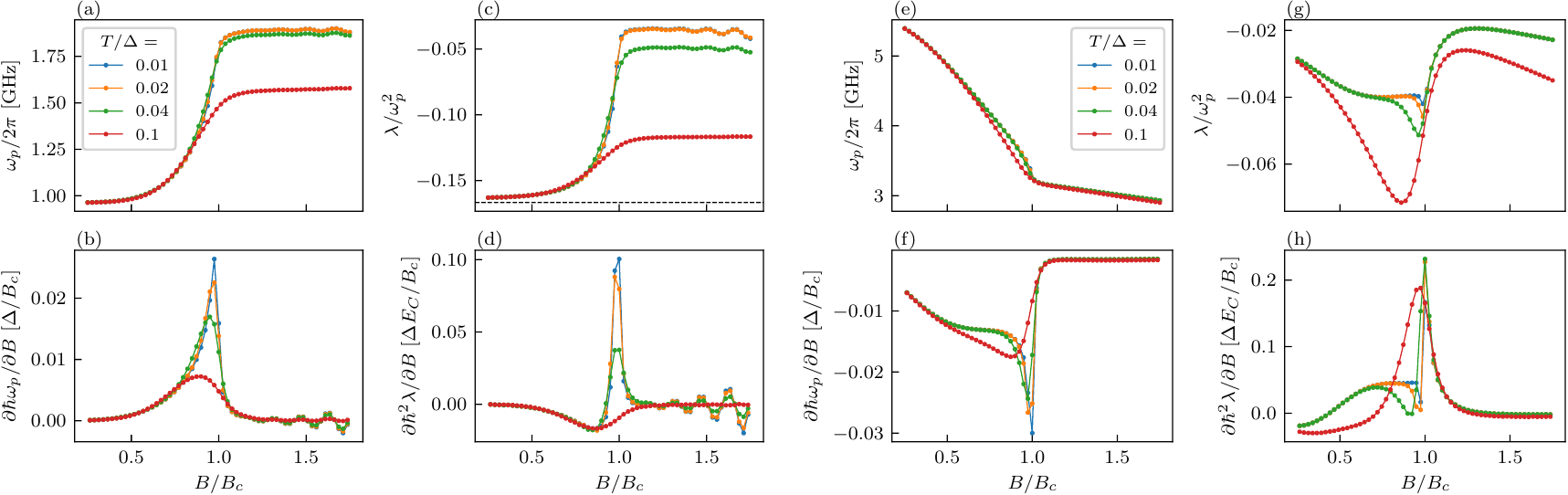}
\caption{
Dependence of the Josephson plasmon frequency $\omega_p$ and the anharmonicity parameter $\lambda$ on the Zeeman energy $B$, for a short junction at finite temperature. Parameter values are the same as in Fig.~\ref{fig:B1}, with $\mu = 3\Delta$. The derivatives of $\omega_p$ and $\lambda$ with respect to $B$ are also shown as a function of $B$. In the thermodynamic limit and at zero temperature, the topological phase transition takes place at $B_c=\sqrt{\Delta^2+\mu^2}$, $B>B_c$ corresponding to the topological regime.
(a-d) "Tunneling" regime ($\gamma=0.6$), (e-h) "Transparent" regime ($\gamma=1$).
  }
 \label{fig:B1T}
\end{figure*}

Thus far, we have concentrated on short junctions either in the tunneling or in the transparent regimes. In junctions of intermediate transparency, $\omega_p$ does not display a kink or a sharp upturn at the topological phase transition; the peak of $\partial\omega_p/\partial B$ at $B=B_c$ can then be less pronounced. Even in those cases, we still observe a clear feature in $\partial\lambda/\partial B$ at $B=B_c$.

Longer junctions show the same phenomenology as the short junctions when it comes to the behavior of $\omega_p$ and $\lambda$ at the topological phase transition (we have simulated weak links with up to $50$ sites). 
However, one additional aspect of the longer junctions is that they may host accidental (imposed neither by electronic topology nor by symmetry), parameter-sensitive, zero-energy touchings at $\delta=0$ in the topologically {\em trivial} phase\cite{vuik2019}. 
Because of particle-hole symmetry inherent to the single-particle spectrum, the positive-energy and negative-energy Andreev states that cross at zero energy have opposite curvatures and anharmonicities.
As a result, the curvature of the {\em occupied} (negative-energy) single-particle states changes abruptly as a function of $B$ at the accidental zero-energy touchings.  A corresponding discontinuity takes place in $\omega_p$ and $\lambda$ (see Fig.~\ref{fig:long_disc} for an example).
The magnitude of the discontinuities is diminished when the $\delta$-dependence of the ABS at the zero-energy touching is smaller; hence, the effect is generally less pronounced in tunnel junctions.
We note that no such discontinuities occur at the true topological phase transition, where positive- and negative-energy Andreev bound states coalesce at zero energy (as opposed to crossing it) as a function of $B$.
Hence, measuring $\omega_p$ and $\lambda$ as a function of $B$ may allow to distinguish an accidental gap closing in the trivial phase (with a discontinuity in $\omega_p$) from a gap closing associated to a topological phase transition (with no discontinuity in $\omega_p$ but a peak in $\partial\omega_p/\partial B$).

Unfortunately, the aforementioned distinction is blurred at finite temperatures, where the discontinuities are transformed into rapid but continuous variations.
Hence, at nonzero temperature, the behavior of $\omega_p$ and $\lambda$ at accidental zero-energy touchings in the trivial phase can mimic the signatures of the topological phase transition. 
In this case, one way to disentangle the true MBS signatures is to analyze the dependence of the peaks in $\mu$; experiments of this type are feasible\cite{frolov2017}. The MBS-related peak at the phase transition will be displaced in $B$ as $B_c = \sqrt{\mu^2+\Delta^2}$, while the peaks of trivial origin will not generically show such dependence, and can even disappear completely as $\mu$ is varied (see Fig.~\ref{fig:long_disc}).

To conclude this subsection, we remark that our main results do not rely on the fact that a magnetic field has been used to induce a topological phase transition. 
Similar conclusions are reached when $B$ is held fixed and the topological phase transition is driven by variations in $\mu$.
In that case,
we find that $\partial\omega_p/\partial\mu$ and $\partial\lambda/\partial\mu$ show strong peaks at the topological phase transition, for the same reasons as the ones alluded to above.

\subsection{Plasmon frequency and anharmonicity parameter at finite temperature}\label{ssec:wpFinitT}

In the preceding subsection, we have investigated $\omega_p$ and $\lambda$ at zero temperature, and have identified signatures of the topological phase transition therein. 
In this subsection, we study the effect of finite temperature on those features. 
We concentrate on temperatures that are low compared to the proximity-induced $s$-wave superconducting gap $\Delta$.
We find that the peaks of $\partial\omega_p/\partial B$ and $\partial\lambda/\partial B$ at the topological phase transition are thermally broadened, though they remain noticeable at experimentally attainable temperatures.
We also find a nontrivial temperature dependence of $\omega_p$ in the topological phase of tunnel junctions, which can be ascribed to the MBS localized on the opposite extremities of the weak link.

Our results are summarized in Fig.~\ref{fig:B1T}. The main aspects of this figure can be understood from the finite temperature expressions (derived from Eqs.~(\ref{eq:UT}) and (\ref{eq:jba_par})) for the plasmon frequency and the anharmonicity parameter near $\delta=0$,
\begin{align}
\label{eq:omegaT}
    \omega_p^2 &= 
    -\frac{4 E_C}{\hbar^2} 
    \sum_{n>0} \tanh\left[\frac{\epsilon_n(0)}{2 k_B T}\right] \epsilon_n^{(2)}(0)\nonumber\\
    \lambda &=-\frac{2 E_C}{3 \hbar^2} \sum_{n>0}\left[\tanh\left[\frac{\epsilon_n(0)}{2 k_B T}\right] \epsilon_n^{(4)}(0)\right.\nonumber\\
    &\left.+\frac{3}{2 k_B T}{\rm sech}^2\left[\frac{\epsilon_n(0)}{2 k_B T}\right]\left(\epsilon_n^{(2)}(0)\right)^2\right],
\end{align}
where $\epsilon_n(0)$ is the energy of the $n$-th single particle state at $\delta=0$ (recall that $\epsilon_n(\delta)\geq 0$ for $n>0$), and $\epsilon_n^{(k)}(0)$ denotes the $k$-th derivative of $\epsilon_n$ with respect to $\delta$ (evaluated at $\delta=0$).

It follows from Eq.~(\ref{eq:omegaT}) that, whenever the energy of the lowest-lying $n>0$ state at $\delta=0$ largely exceeds $k_B T$, $\omega_p$ and $\lambda$ will have a negligible temperature-dependence. In other words, the condition for a significant temperature-dependence of $\omega_p$ and $\lambda$ is that
$k_B T\gtrsim \epsilon_1(0)$  (with $\epsilon_1(0)$ the smallest single-particle eigenenergy)
Another useful piece of information in Eq.~(\ref{eq:omegaT}) is that, for a fixed $n$, the contribution of the given single-particle level to $\omega_p$ is suppressed when $k_B T \gg \epsilon_{n}(0)$. 

Let us now analyze in detail the content of Fig.~\ref{fig:B1T}. We divide the discussion in three parts, according to three regimes: (i) the trivial phase, (ii) the onset of the topological phase, and (iii) deep in the topological phase.

In the trivial phase of short junctions, there are no subgap states in our model at $\delta=0$ (regardless of the transparency) \cite{vuik2019}.
Therefore, 
$\epsilon_{1}(0)=\Delta_-$ 
and the temperature-dependence of $\omega_p$ and $\lambda$ is strong only if $k_B T\gtrsim\Delta_-$. 
At low temperatures, such condition is satisfied only close to the topological phase transition (because $\Delta_-\to 0$ when $B\to B_c$). This explains why, in Figs.~\ref{fig:B1T}a, c, e and g, the curves for different temperatures are superimposed deep in the trivial phase, but begin to diverge closer to the topological phase transition.
The fact that $k_B T \gg \Delta_-$ is easily attained at the topological phase transition also explains why the zero-temperature kinks in $\omega_p$ and $\lambda$, which take place at the topological phase transition in highly transparent junctions (Fig.~\ref{fig:B1}e-h), are washed out by thermal effects. 
Still, remnants of the kinks survive at low ($k_B T\ll \Delta$) but finite temperatures, in the form of broadened peaks for  $\partial\omega_p/\partial B$ and $\partial\lambda/\partial B$.
These peaks are evident in Figs.~\ref{fig:B1T}b, d, f and h.

At the onset of the topological phase ($B\simeq B_c$), four MBS emerge.
Two of them, localized on the outer extremities of the superconducting electrodes, have near-zero energies that are essentially independent of $\delta$ in the vicinity of $\delta=0$; these dispersionless states do not contribute to $\omega_p$ and $\lambda$. For this reason, we will hereafter omit the two "outer" MBS from the discussion.
In contrast, the two "inner" MBS localized on the opposite extremities of the weak link have a non-negligible bandwidth in $\delta$, given by the hybridization energy $E_M$; these dispersive states can contribute appreciably to $\omega_p$ and $\lambda$.
In a junction of high transparency, 
$\epsilon_{1}(0)=\Delta_-\ll E_M\simeq \Delta_+$.
Hence, in such junction, the leading temperature-dependence of $\omega_p$ and $\lambda$ at $k_B T\gtrsim \Delta_-$ originates mainly from the non-Majorana states.
In contrast, in a junction of low transparency, 
$\epsilon_{1}(0)=E_M<\Delta_-\ll \Delta_+$ 
and the leading temperature-dependence at $k_B T \gtrsim E_M$ originates from the inner MBS.

\begin{figure*}[tb]
    \includegraphics[width=0.995\textwidth]{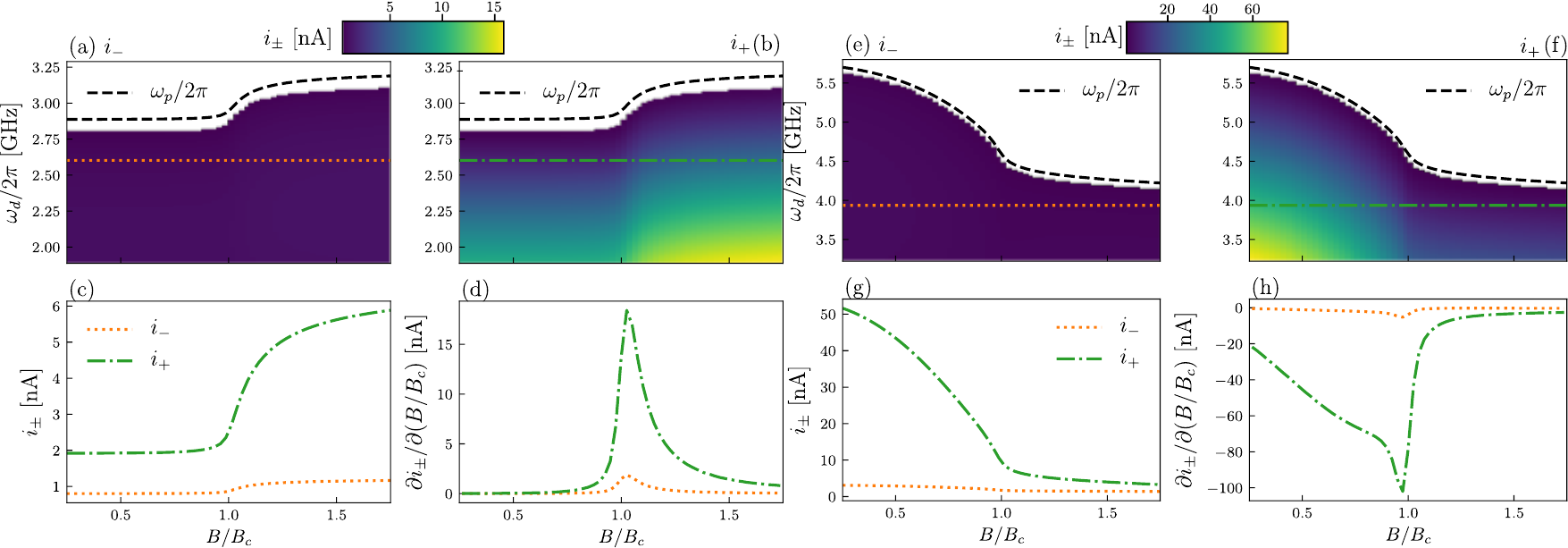}
\caption{(a,b) Behavior of lower and upper bifurcation currents -- respectively $i_-$ in panel (a) and $i_+$ in panel (b) -- in a short driven Josephson junction with $\gamma=0.6$, as a function of the Zeeman energy $B$ and drive frequency $\omega_d$. 
The dashed black line indicates the plasmon frequency in GHz ($\omega_p/2\pi$). The white regions on the top panels denote areas of parameter space where no bifurcation arises. 
(c) Line cuts of panels (a,b) at fixed drive frequency $\omega_d/2\pi =2.6$~GHz. The corresponding cuts are indicated in panel (a) (dotted orange horizontal line) and panel (b) (dashed-dotted green horizontal line).
(d) The derivatives $\partial i_\pm/\partial B$ display peaks at the topological phase transition ($B=B_c$). These peaks are gradually broadened as temperature is increased (data not shown).
(e-h) Same as panels (a-d) but for the case of a transparent junction where $\gamma=1$. Line cuts in panel (g) correspond to $\omega_d/2\pi \approx 3.94$~GHz.
In all panels, the parameters are $T=0.02 \Delta/k_B$, $\mu = \Delta$, $E_C/h=0.1$ GHz, and we add an external Josephson energy $E_{J0}/h = 10$ GHz.
 }
 \label{fig:bif1}
\end{figure*}

As the system is driven deeper into the topological phase, Fig.~\ref{fig:B1T} reflects two possible scenarios. 
In the scenario of Fig.~\ref{fig:B1T}e-h, the junction is highly transparent and there are no dispersive subgap states at $\delta=0$.
In this case, 
$\epsilon_1(0)={\rm min}(\Delta_-,\Delta_+)$, where we once again ignore the outer MBS.
If $\Delta_-< \Delta_+$ (which is common not far from the topological phase transition), 
$\epsilon_1(0)=\Delta_-$ grows linearly with the magnetic field.
Therefore, the temperature-dependence of $\omega_p$ and $\lambda$ weakens gradually as $B$ grows, and becomes once again negligible when ${\rm min}(\Delta_-,\Delta_+)\gg k_B T$. This is why, in Fig.~\ref{fig:B1T}e and g, the curves for different temperatures tend to converge deep in the topological phase.

Contrastingly, in the scenario of Fig.~\ref{fig:B1T}a-d, the junction has low transparency. In this case, there are dispersive subgap MBS of energy $E_M$ at $\delta=0$, so that 
$\epsilon_1(0)=E_M$. 
Moreover, the low junction transparency implies $E_M \ll \Delta_-$. 
Then, as temperature is raised well above $E_M$, the contribution of MBS to $\omega_p$ is suppressed via the $\tanh(E_M/2 k_B T)\ll 1$ factor in Eq.~(\ref{eq:omegaT}).
At the same time, the energies of the higher-energy (non-Majorana) scattering states are relatively dispersionless in $\delta$ and therefore make a relatively modest contribution to $\omega_p$, much like in the trivial phase. 
As a result, $\omega_p$ is strongly reduced in the topological phase when $k_B T \gg E_M$.
This explains the traits of Fig.~\ref{fig:B1T}a in the topological phase. 

One intriguing outcome from the foregoing discussion is that it appears to be possible to thermally switch on and off the contribution from MBS to $\omega_p$ in short tunnel junctions, by going from $k_B T\ll E_M$ to $k_B T\gg E_M$. 
This statement applies to longer junctions of low transparency as well.
Although such junctions host low-energy Andreev-bound states in the trivial phase, 
their $\delta$-dependence in the tunneling regime is relatively small; therefore, changing their thermal occupancy does not make a substantial difference to $\omega_p$. 
We have verified this point for junctions with 20 and 50 sites in the weak link, and $|\mu|\ll \Delta$.

\subsection{Bifurcation currents}\label{ssec:bif}

Thus far, we have been concerned with the prediction and understanding of peaks in the derivatives of $\omega_p$ and $\lambda$ across a topological phase transition. 
The connection between $\omega_p$ and $\lambda$ with the operation of a JBA is made by Eq.~(\ref{eq:i_pm}). 
From this equation, it is clear that the bifurcation currents of the TJBA will display signatures of a topological phase transition. 

The results are summarized in Fig.~\ref{fig:bif1}, which shows the lower and upper bifurcation currents as a function of the magnetic field, for short junctions of low (Fig.~\ref{fig:bif1}a-d) and high (Fig.~\ref{fig:bif1}e-h) transparency. 
The drive frequency $\omega_d$ is chosen to be below $\omega_p$, as otherwise there is no bifurcation (cf. Sec.~\ref{sec:TJBA}). 

The bifurcation currents are much smaller in the tunneling regime than in the transparent regime, because the critical current of the junction is much lower in the former.
At the topological phase transition, the behaviour of bifurcation currents as function of magnetic field parallels that of $\omega_p$, and thus their derivatives display peaks at $B=B_c$. These peaks are broadened at finite temperature.

\section{Detection of the $4\pi$ Josephson effect}
\label{sec:4pi}

One of the characteristic signatures of the presence of MBS in a Josephson junction is the $4\pi$-periodic Josephson effect\cite{kitaev2001,fu2008}.
The detection of this effect has been an active research topic in recent years. 
The leading experimental reports include missing Shapiro steps\cite{rokhinson2012, bocquillon2016} and the halving of the Josephson radiation frequency\cite{laroche2019}.
In this section, we describe an alternative (albeit indirect) microwave signature of the $4\pi$ Josephson effect in short tunnel junctions, which is amenable to detection by a TJBA.

\begin{figure}[t]
    \includegraphics[width=0.99\columnwidth]{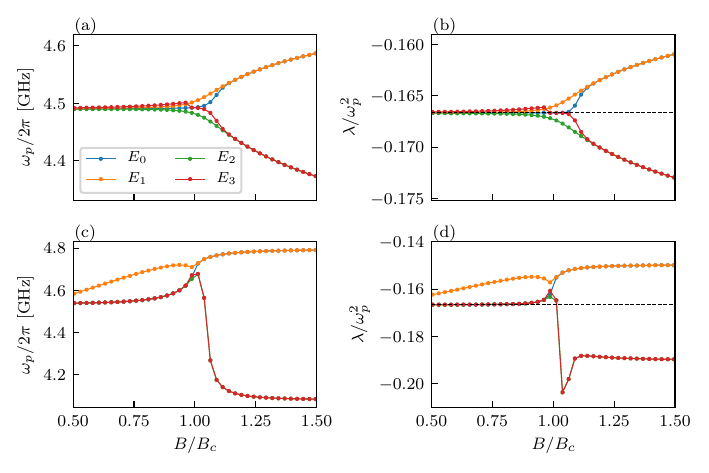}
\caption{Representative examples of the plasmon frequency $\omega_p$ and the anharmonicity parameter $\lambda$ near $\delta=0$ for the four lowest-lying many-body states in a short, tunnel ($\gamma=0.6$) junction. The junction is shunted with an ordinary Josephson junction with Josephson energy $E_{J0}/h = 25$~GHz. The charging energy is $E_C/h=100$~MHz. The two rows correspond to two different values of the chemical potential with (a,b) $\mu=0$, (c,d) $\mu=2\Delta$. The transparency of the junction is larger for $\mu=2\Delta$. In the trivial phase of a junction with low transparency (panels (a,b)), $\omega_p$ and $\lambda$ are all similar to one another.
At the onset of the topological phase, 
$\omega_p$ and $\lambda$ split into two different values. For sufficiently small values of $E_{J0}$, the lower branch of $\omega_p$ and $\lambda$ disappears when $B$ exceeds a critical magnetic field (this critical field is larger than $B_c$).}
 \label{fig:many_omega}
\end{figure}

We begin by showing $\omega_p$ and $\lambda$ in a short tunnel junction, as a function of the magnetic field, for several of the lowest-energy many-body states (Fig. \ref{fig:many_omega}).
In order to obtain these results, $U(\delta)$ in Eq.~(\ref{eq:jba_par}) has been replaced by the energy of individual many-body states. 
This is unlike in Sec.~\ref{sec:TPT}, where we concentrated on $\omega_p$ and $\lambda$ in thermodynamic equilibrium.
In the lowest-energy many-body state, all single-particle energy levels $n$ with $\epsilon_n(\delta)<0$ are occupied; its corresponding energy is given by Eq.~(\ref{eq:U_T0}). 
The excited many-body states are constructed by creating particle-hole excitations on top of the lowest-energy state (one such excitation involves emptying a single-particle state $-n$ and populating a single-particle state $n$, for $n>0$); their corresponding energies are given by adding to Eq.~(\ref{eq:U_T0}) the excitation energies of the particle-hole pairs. 

According to Fig. \ref{fig:many_omega}a and b, in the trivial phase of a tunnel junction, $\omega_p$ and $\lambda$ are very similar for all the low-lying many-body states. 
Then, at the onset of the topological phase, both $\omega_p$ and $\lambda$ split into two branches, which diverge gradually from one another as the magnetic field increases. Deeper in the topological phase, the lower branch of $\omega_p$ can reach zero at a critical value of the magnetic field and subsequently disappear.
The magnitude of this upper critical field depends on microscopic details, and can be partly controlled by shunting the topological Josephson junction with an ordinary Josephson junction of Josephson energy $E_{J0}$ (with $E_{J0}>0$). As mentioned in Sec.~\ref{ssec:wpT0}, $E_{J0}$ may also model the contribution of multiple subbands to the Josephson energy in a quasi one-dimensional nanowire. The larger the value of $E_{J0}$, the higher the critical field at which the lower plasmon branch disappears in the topological phase. For large enough $E_{J0}$, the frequency of the lower plasmon branch saturates before reaching zero and the branch never disappears.

\begin{figure}[t]
    \includegraphics[width=\columnwidth]{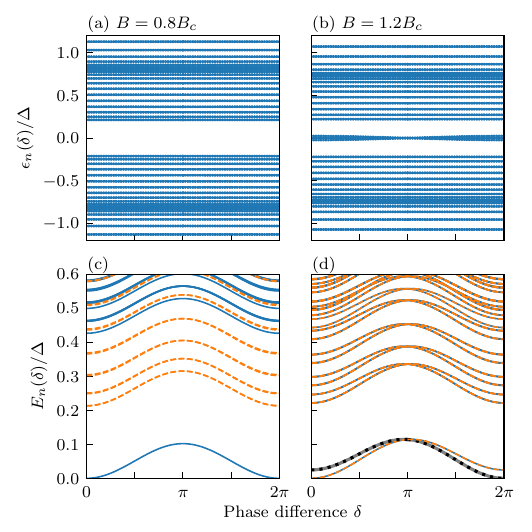}
\caption{(a) A representative example of the single-particle spectrum in a short tunnel junction, in the topologically trivial phase ($B=4B_c/5$, $\mu=0$). The single-particle states are dispersionless and there are no subgap states. (b) A representative example of the single-particle spectrum in a short tunnel junction, in the topological phase ($B=6B_c/5$, $\mu=0$). The Majorana bound states are visible inside the bulk superconducting gap. The dispersion of the inner MBS is strong when compared to those of the scattering states. (c) and (d) Low-lying many-body states built from (a) and (b), respectively. In the case of many-body states, we have added a term $E_{J0}(1-\cos\delta)$ to all states, with $E_{J0}/\Delta = 0.05$; this energy originates from an ordinary Josephson junction connected in parallel. In the trivial phase, all many-body states have similar shapes, dictated by $-E_{J0} \cos\delta$. In the topological phase, a Majorana term $-(1)^p E_M \cos(\delta/2)$ emerges, where $(-1)^p$ is the parity of the occupation of the inner MBS.
As a consequence, the first many-body excited state, corresponding to a parity-flip of the inner MBS, has a metastable minimum at $\delta=0$ (thick solid grey and dotted black curves). 
The solid blue (dashed orange) curves corresponds to even (odd) global fermionic parity states.
} 
 \label{fig:spectra}
\end{figure}

The preceding observations from Fig.~\ref{fig:many_omega}a and b are closely related to the emergence of the $4\pi$-periodic Josephson effect in the topological phase. Next, we explain this connection by analyzing the energy spectrum of the junction (Fig.~\ref{fig:spectra}). 

In the trivial regime of the tunnel junction, the dispersion of the single-particle states as a function of $\delta$ is approximately flat (Fig.~\ref{fig:spectra}a).
Hence, when $E_{J0}$ is sizeable, the energies of the many-body states (Fig.~\ref{fig:spectra}c) are of the form $-E_J \cos\delta+{\rm const}$, where $E_J\simeq E_{J0}$ and "const" is a term independent of $\delta$ (though it takes different values for different many-body states).
Note that $E_J$ is approximately the same for all the many-body states. 
This explains why $\omega_p$ and $\lambda$ are very similar for different many-body states in the topologically trivial regime.

The situation is different in the topologically nontrivial regime of the tunnel junction, where the subgap single-particle states associated to the inner MBS display increased dispersion (Fig. \ref{fig:spectra}b); this dispersion is qualitatively of the form $\pm E_M \cos(\delta/2)$, where $E_M>0$ is the hybridization energy of the inner MBS.

The occupation of these dispersive Majorana bands determines the parity $p$ of the inner MBS.
Many-body states have a well-defined value of $p$, with $p=0$ ($1$) if the localized single-particle level due to the hybridized MBS is empty (filled).
Then, the energies of the many-body states take the approximate form $-E_J \cos\delta -(-1)^p E_M \cos(\delta/2) +{\rm const}$, where $E_J\simeq E_{J0}$. 
Once again, the magnitude of $E_J$ and $E_M$ does not change from one many-body state to another.
The $\cos(\delta/2)$ term in the energy is responsible for the $4\pi$-periodic Josephson effect; indeed, if $p$ is conserved, the many-body energies (and thus the supercurrent) are $4\pi$-periodic in $\delta$.

We now come to the key observation. In the topological phase, the many-body states with $p=0$ have an absolute minimum at $\delta=0$ (mod $4\pi$) (see Fig. \ref{fig:spectra}d).
In contrast, the many-body states with $p=1$ show a {\em local} (metastable) minimum at $\delta=0$ (mod $4\pi$), if and only if $E_M\in(0,4 E_J)$.
This local minimum exists because of the coexistence of $-E_J\cos(\delta)$ and $+E_M\cos(\delta/2)$ terms in the energy.
When it exists, the local minimum of $p=1$ states at $\delta=0$ differs in curvature and anharmonicity from the absolute minima of the $p=0$ states at $\delta=0$.
As a result, from Eq.~(\ref{eq:jba_par}), the plasmon frequencies and anharmonicities associated to small oscillations around $\delta=0$ are $p$-dependent:
\begin{align}
  \hbar\omega_p (p) &= \sqrt{8 E_C \left(E_J+(-1)^p\frac{E_M}{4}\right)}\nonumber\\
  \lambda (p) &= -\frac{4 E_C}{3 \hbar^2} \left(E_J+(-1)^p\frac{E_M}{16}\right).
\end{align}
When $E_M=0$, $\omega_p$ and $\lambda$ are identical for all many-body states.
As the system enters the topological phase, $E_M$ grows together with $\Delta_-$ and two distinct values of $\omega_p$ and $\lambda$ emerge for $p=0,1$.
This explains the main trends observed in Fig.~\ref{fig:many_omega}a and b.

If $E_M=4 E_J$, the curvature of the local minimum for $p=1$ states at $\delta=0$ vanishes, and thus $\omega_p(1)=0$.
For $E_M>4 E_J$, the many-body states with $p=1$ have a maximum (rather than a local mimimum) at $\delta=0$; small oscillations can only take place in the vicinity of $\delta=2\pi$. But, because the $\delta=2\pi$ minimum for $p=1$ states is identical to the $\delta=0$ minimum for $p=0$ states, it follows that there is only one branch of $\omega_p$ and $\lambda$ when $E_J>4 E_M$.

When $E_{J0}\simeq 0$, a magnetic field slightly higher than the one required to attain the topological phase is sufficient to lead to $E_M > 4 E_J$. The reason for this is that 
$E_M/E_J$, being inversely proportional to $\gamma$, can be large in the tunneling regime (at least for the single-channel nanowire we consider). 
The interval of the magnetic field in which the metastable minimum is present is enhanced by increasing $E_{J0}$. 
For a sufficiently large $E_{J0}$, $E_M$ never exceeds $4 E_{J0}$, no matter how high $B$ is; an example of this situation is illustrated in Fig.~\ref{fig:many_omega}, where the lower branch of $\omega_p$ saturates at a finite value at high $B$. This saturation has to do with the fact that $\Delta_+$, decreasing with $B$, eventually becomes smaller than $\Delta_-$ deep in the topological phase, thereby stunting and reversing the growth of $E_M$ as the magnetic field increases.

\begin{figure}[t]
\includegraphics[width=1.0\columnwidth]{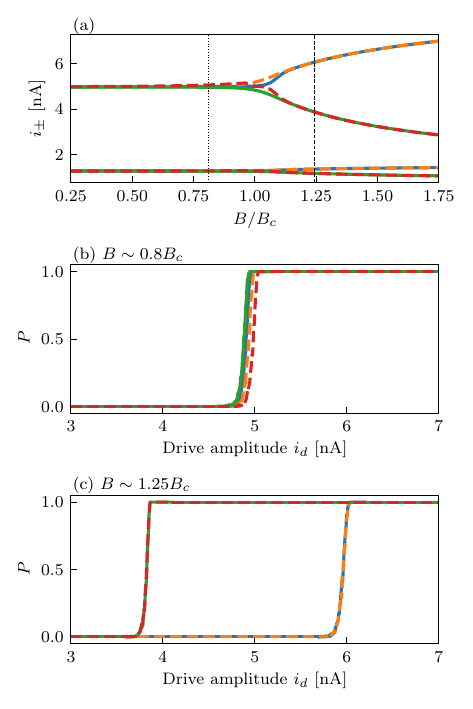}
\caption{(a) Lower and upper bifurcation currents for the first four many-body states, in a short tunnel junction ($\gamma=0.6$, $\mu=0$; see Fig.~\ref{fig:many_omega} for other parameters). The bifurcation currents for oscillations about $\delta=0$ are split into two branches in the topological phase, depending on the parity $(-1)^p$ of the inner Majorana bound state ($p=0$ for orange and blue curves, $p=1$ for green and red curves).
(b,c) Bifurcation probability $P$ as a function of the RF-drive amplitude for the four lowest energy many-body states. (b) $B\sim 0.9 B_c$ and (c) $B \sim 1.25 B_c$. The variance of the current noise is chosen as $\sigma_N=2$~nA.
 }
 \label{fig:many_bif}
\end{figure}

The aforementioned features are echoed in the bifurcation currents of the TJBA.
Figure~\ref{fig:many_bif}a shows $i_+$ and $i_-$ for the lowest many-body states, in a short tunnel junction.
In the trivial phase, the bifurcation currents for the low-lying many-body states are all similar. Beyond the topological phase transition, $i_+$ and $i_-$ split into two branches corresponding to $p=0$ and $p=1$ states.
Accordingly, the switching probabilities between low- and high-amplitude oscillating states near $\delta=0$ become $p$-dependent (see Fig.~\ref{fig:many_bif}b-c).
Roughly, $p=0,1$ can be regarded as two quantum states with distinct JBA switching probabilities.

Let us now discuss the experimental observability of the 4$\pi$-periodic Josephson effect in a TJBA.
The underlying idea is to induce transitions between the absolute and local minima of the energy spectrum, and to measure the frequency of small oscillations of $\delta$ around those minima quickly enough (before energy relaxation to the absolute minimum takes place). 
We begin by noting that the vertical-in-$\delta$ transition between the $p=0$ and $p=1$ minima cannot be induced by microwave pulses, because electron-photon interactions preserve the MBS parity\cite{lopes2019}.
This is unlike the case of the two quantum states in a superconducting qubit, the transition between which is routinely induced by microwave pulses.

Nonetheless, there are still two ways one can probe the presence of the metastable minimum with a JBA scheme.
One is by letting the system spontaneously switch between the stable and metastable minima.
This will occur due to quasiparticle poisoning or other vertical-in-$\delta$ processes that violate the conservation of $p$.
By repeating a JBA measurement many times for a fixed $B$ and $\mu$, one will find two very different switching probabilities on the topological phase (provided that $E_M\gg k_B T$), but not on the trivial phase.
In addition, this type of measurement may allow to  extract the lifetime $\tau_p$ of the inner MBS parity $p$ experimentally (provided that the escape time of $\delta$ away from the local minimum through thermal and quantum fluctuations is longer than $\tau_p$).

A more deterministic way of switching between stable and metastable minima is by applying current pulses\cite{likharev1986}.  A current pulse of short duration and amplitude comparable to the critical current of the junction can be applied to vary $\delta$ away from one minimum of the energy. For fixed $p$, changes in $\delta$ that are odd multiples of $2\pi$ will connect a local minimum with an absolute minimum (see Fig.~\ref{fig:spectra}d). The precise control of $\delta$ with the current pulse is not required, however; what is important is that, after the current pulse stops, $\delta$ will begin to oscillate around the nearest minimum. If $E_M\in(0, 4 E_J)$, sometimes this minimum will be a local minimum, and other times it will be an absolute minimum.
By measuring the state of the oscillator (low-amplitude vs high-amplitude) immediately after the current pulse is switched off, and repeating the whole process multiple times, two distinct switching probabilities will be apparent in the topological phase, but not in the trivial phase. 
This measurement protocol is analogous to the JBA-based measurement of Rabi oscillations and relaxation times in conventional superconducting qubits\cite{vijay}.

For the preceding measurement protocol to function, it is necessary that 
the lifetime of the metastable minimum be long compared to the JBA measurement time, the latter being of the order of $0.1\mathit{-}1$ microseconds\cite{vijay}.
As mentioned above, the metastable minimum has a lifetime that is in part determined by the quantum/thermal fluctuations over the local potential barrier, as well as vertical-in-$\delta$ inelastic processes that flip $p$.
When $4 E_J - E_M \gg k_B T$, thermal fluctuations over the barrier are exponentially suppressed. Large quantum fluctuations of the phase (tunneling across the energy barrier around the metastable minimum) are likewise exponentially suppressed if the charging energy of the junction is small compared to the Josephson energy.
Under these conditions, the lifetime of the metastable minimum is limited by quasiparticle poisoning processes, which are indeed slow compared to the timescale in which a JBA experiment can be realized\cite{hays2018}.

So far, we have restricted our discussion to short tunnel junctions.
The reason for this is that, in more transparent junctions of any length, the situation is less favorable for the TJBA-based detection of the $4\pi$-periodic Josephson effect. This is so for two main reasons.
First, the scattering states display significant dispersion in more transparent junctions.
Therefore, there is a wide spread of $\omega_p$ and $\lambda$ for different many-body states, even in the trivial regime; this behavior is already incipient in Fig.~\ref{fig:many_omega}c and d, for junctions with $\gamma=0.6$.
Second, for $B\gtrsim B_c$, the Majorana bands near $\delta=0$ are buried within the scattering states (because $E_M\gg \Delta_-$ in a transparent junction).
As a result, the many-body states with $p=1$ are part of a continuum of states, which drastically decreases their lifetime and observability. 

Concerning longer tunnel junctions (up to 50 sites in the weak link were studied), they do show the desired effect, but it is masked by the fact that there abound non-Majorana subgap states that disperse significantly.  This leads to an array of different $\omega_p$ and $\lambda$ for the low-lying many-body states, in the topological phase. Yet, the fact that one of the plasmon frequencies drops appreciably in the vicinity of the topological phase transition (even vanishing at a critical magnetic field if $E_{J0}$ is sufficiently small) still holds and is related to the emergence of MBS.

We conclude this section by comparing our results to earlier works in the literature. 
First, the TJBA-based detection scheme for the $4\pi$-periodic Josephson effect relies on the dynamics of the superconducting phase difference near a minimum of the energy (e.g. $\delta=0$).
As such, it is robust under energy gaps that can occur at $\delta=\pi$ due to parity violating processes. It also differs from existing proposals and experiments to measure the $4\pi$ Josephson effect. The latter rely on swiping $\delta$ over multiple full periods along a single many-body state of fixed $p$; the swipe must be fast enough to overcome any putative energy gap at $\delta=\pi$ (through Landau-Zener tunneling).

Second, the splitting of $\omega_p$ and $\lambda$ that we have reported is related to features previously predicted for microwave spectroscopy \cite{ginossar2014, keselman2019, avila2020}. 
In Ref.~[\onlinecite{ginossar2014}], it was found that energy levels of a usual transmon qubit are split into doublets when $E_M\neq 0$. 
From Fig. 2 in that paper, it is evident that the plasmon frequencies would also be split in two for different sectors of the inner MBS parity. 
Yet, the authors restricted themselves to the regime $E_M\ll E_J$, in which the splitting in $\omega_p$ and $\lambda$ is very small, and also no systematic study of $\omega_p$ and $\lambda$ was carried out as a function of the microscopic parameters of the system. 
Similarly, in Refs.~[\onlinecite{keselman2019}] and~[\onlinecite{avila2020}], a splitting of a spectral line in the microwave spectrum was noted at the topological phase transition in a way that is reminiscent of Fig.~\ref{fig:many_omega}c. These studies focused on the charge qubit parameter regime where the observed splitting was dominated by the charging energy, while our work focus on the complementary transmon parameter regime where the splitting is dominated by $E_M$.
The possibility that the lower branch of $\omega_p$ and $\lambda$ may disappear in the topological phase of tunnel junctions (for $E_M> 4 E_J$) was not recognized in earlier works.

\section{Conclusions}
\label{sec:conclusion}

In summary, we have proposed that a bifurcation amplifier built from a topological Josephson junction (a topological Josephson bifurcation amplifier, or TJBA) can be an interesting device to study various aspects related to Majorana bound states and their emergence in semiconductor nanowires with proximity-induced superconductivity.

Much of our effort has been devoted towards understanding the behavior of the Josephson plasmon frequency $\omega_p$ and the anharmonicity parameter $\lambda$ as a function of the externally applied Zeeman field $B$, in the transmon regime, at zero and finite temperature, as well as for different many-body states. The reason for this focus is that $\omega_p$ and $\lambda$ are key variables that govern the operation of the bifurcation amplifier; small changes in those quantities can be sensed when the TJBA is driven in the vicinity of the bifurcation point. Though our analysis is based on a minimal model of a nanowire (single-channel, free from disorder), we have attempted to extract physical statements and results whose relevance should transcend the model's simplicity.

Our work contains two main results. First, in the regimes of low and high transparency, $\partial\omega_p/\partial B$ and $\partial\lambda/\partial B$ display pronounced peaks at the topological phase transition, which are reflected in the bifurcation currents of the TJBA.
These peaks are associated to the emergence of MBS.
Finite temperature broadens the peaks, which nonetheless remain significant provided that the thermal energy is well below ($< 10\%$) the proximity-induced $s$-wave superconducting gap.
Remarkably, in tunnel junctions, the contribution from MBS to $\omega_p$ can be turned on and off by varying the temperature. 
Another remarkable aspect is the evolution of the ratio $\lambda/\omega_p^2$ across a topological phase transition in a junction of low transparency, at zero temperature. In the trivial phase, $\lambda/\omega_p^2$ tends to a plateau of $-1/6$, which corresponds to the value found in a conventional Josephson junction. 
In contrast, in the topological phase, $\lambda/\omega_p^2$ tends to a plateau of $-1/24$. The $1/4$ factor between the two plateaux originates from a fundamental change in the dependence of the ground state energy on the superconducting phase difference $\delta$: it is $\cos\delta$ in the trivial regime and $|\cos(\delta/2)|$ in the topological regime (the absolute value applies in the absence of fermion parity constraints).
The change is again due to the appearance of MBS.

Second, the TJBA can spot the emergence of unconventional periodicities in the current-phase characteristics of the junction, such as the $4\pi$-periodic Josephson effect that is induced by $B$ when the junction undergoes a topological phase transition.
The TJBA-based approach for measuring the $4\pi$-periodic Josephson effect is centered on the dynamics of $\delta$ near an energy minimum; it therefore differs from proposals and experiments that  rely on swiping $\delta$ over multiple full periods.
The underlying concept relies on monitoring $\omega_p$ and $\lambda$ for different many-body states near $\delta = 0$.
Because of the coexistence between $\cos\delta$ and $\cos(\delta/2)$ terms in the ground state energy, $\omega_p$ and $\lambda$ split into two branches each in the topological phase; the same behavior is present in the bifurcation currents of the TJBA.
This splitting originates from MBS.

The above predictions and results are easier to validate in junctions that contain no subgap states other than the MBS. In our model, such is the case when the junction is short. In longer junctions, zero-energy Andreev bound states arise even in the trivial phase, thereby mimicking the signatures of MBS. In their presence, additional analysis is necessary in order to discern topological signatures from trivial ones. Similar issues complicate most detection schemes of MBS in semiconductor nanowires.

Although a TJBA has not yet been built, the ongoing progress in combining nanowire-based superconducting qubit technology with the magnetic fields required to realize MBS suggests that its fabrication is within reach.
Ideally, the present work may provide an incentive for the experimental development of a TJBA, which will in turn encourage further theoretical studies on the device. Obvious tasks on the theory side include the adoption of more realistic models of nanowires, as well as a fully quantum mechanical treatment of the TJBA dynamics.

{\em Note added:} 
In the final stages of this work, a preprint appeared~\cite{kurilovich2021} in which an analytical theory for the dynamics of a topological Josephson junction coupled to a cavity is presented. The findings of Ref.~[\onlinecite{kurilovich2021}], which include a fork-like feature of the cavity frequency pull at the topological phase transition, are consistent with our results.

\acknowledgements
This research was financed in part by the Canada First Research Excellence Fund and the Natural Science and Engineering Council of Canada. Numerical calculations were partly done with computer resources from Calcul Qu\'ebec and Compute Canada. U.C.M. acknowledges the support from CNPq-Brazil (Project No. 309171/2019-9). 

\section*{Data availability}
The data that support the findings of this study are available from the corresponding author upon reasonable request.




\begin{thebibliography}{100}
\bibitem{tunnel1} While the Josephson energy of the junction is reduced in the tunneling regime, we still assume that it is large compared to the charging energy of the junction.
\bibitem{EJ0} The implicit assumption here is that $E_{J0}$ will have a  relatively featureless (or at least smooth) dependence in $B$, for the values of $B$ that are required to induce a topological phase transition.
\bibitem{tunnel} This discussion does not apply to junctions of low transparency, because scattering at the interface between the electrodes and the weak link invalidates the picture of two quasi-independent $p-$wave gaps.
\end{thebibliography}
\end{document}